\documentclass[3p]{elsarticle}

\usepackage{epstopdf}
\usepackage{amsmath}
\usepackage[psamsfonts]{amsfonts}
\usepackage{theorem}
\usepackage{version}
\usepackage{graphicx} 
\usepackage{lscape} 
\usepackage{natbib}

\usepackage{color,hyperref}
\definecolor{darkblue}{rgb}{0.0,0.0,0.3}
\hypersetup{colorlinks,breaklinks,
            linkcolor=darkblue,urlcolor=darkblue,
            anchorcolor=darkblue,citecolor=darkblue}

\usepackage{enumerate, theorem}
\usepackage{graphicx} 
\usepackage{version}

\newtheorem{proposition}{Proposition}[section]

\newcommand{\cqfd}{\hfill $\square$}

\newcommand{\R}{\mathbb R}

\newcommand{\N}{\mathbb N}

\newcommand{\n}{^{(n)}}

\newcommand{\thetab}{{\pmb \theta}}

\newcommand{\varthetab}{{\pmb \vartheta}}

\newcommand{\Thetab}{{\pmb \Theta}}

\newcommand{\Deltab}{{\pmb \Delta}}
\newcommand{\taub}{{\pmb \tau}}
\newcommand{\nub}{{\pmb \nu}}

\newcommand{\betab}{{\pmb \beta}}

\newcommand{\pr}{^{\prime}}

\newcommand{\uut}[1] {\renewcommand{\arraystretch}{0.5}
\begin{array}[t]{c}{#1}\vspace{.2mm}\\
   \widetilde{}\vspace{-2.1mm}
\end{array}
\renewcommand{\arraystretch}{1} }

\newcommand{\ut}[1] {\renewcommand{\arraystretch}{0.5}
\begin{array}[t]{c}{#1}\\
   \widetilde{}
\end{array}
\renewcommand{\arraystretch}{1}}
\newcommand{\utQ}{\!\! \uut{Q}\!\!}
\newcommand{\utDelta}{\!\! {\uut {\pmb \Delta}}\vspace{-1mm}\!\!}
\newcommand{\utbeta}{\!\! {\uut {\pmb \beta}}\vspace{-1mm}\!\!}

\newcommand{\DS}[1]{\displaystyle{#1}}
\newcommand{\ny}{n\rightarrow\infty}

\begin{document}
\begin{frontmatter}

\title{One-Step R-Estimation in 
 Linear Models  with 
  Stable Errors}
  
  \author{ Marc {Hallin}$^{\rm a, b}$, Yvik {Swan}$^{\rm c}$,
Thomas Verdebout$^{\rm d}$ and David Veredas$^{\rm a}$\\ 
$^{\rm a}${\em{Institut de Recherche en Statistique, ECARES, and
    ECORE, Universit\' e libre de Bruxelles, Belgium.} }  \\ 
$^{\rm b}${\em{Acad\' emie Royale de Belgique,     ORFE, Princeton
    University, and CentER, Tilburg University.}}  \\ 
 $^{\rm c}${\em{Unit\'e de Recherche en  Math\' ematiques, Universit\' e libre
     du Luxembourg, Luxembourg. }} \\ 
 $^{\rm d}${\em{EQUIPPE-GREMARS,   Universit\' e Lille Nord de France.    
  }}}


%


%

%

\begin{abstract} 
 Classical estimation techniques for linear models either are inconsistent, or perform rather poorly, under~$\alpha$-stable error densities; most of them are not even rate-optimal.      In this paper, we propose an original one-step R-estimation method and investigate its asymptotic performances under stable densities. Contrary to traditional least squares,   the proposed R-estimators   remain root-$n$ consistent (the optimal rate) under the whole family of stable distributions, irrespective of their asymmetry and tail index. While parametric stable-likelihood estimation, due to the absence of a closed form for stable densities, is quite cumbersome, our method allows us to construct estimators reaching the parametric efficiency bounds associated with any prescribed values $(\alpha_0,  \  b_0)$ of the tail index $\alpha$ and skewness parameter $b$, while preserving root-$n$ consistency under any~$(\alpha, \ b )$ as well as under usual light-tailed densities. The method furthermore  avoids all forms of multidimensional argmin computation. Simulations confirm its excellent finite-sample performances. 
\end{abstract}
\end{frontmatter}

\section{Introduction.}\label{intro}

Evidence of heavy-tailed behavior and infinite variances in economics and, even more so, in finance and insurance, is overwhelming. In such context, the Gauss-Markov theorem for linear regression\footnote{Recall that the Gauss-Markov theorem establishes, for errors with finite variance,  that OLS  estimators are  {\it best linear unbiased}  estimators.  } no longer holds  true, and the usual OLS   estimators of regression coefficients lose their theoretical justifications. Much worse: they also lose their traditional\footnote{Under the classical condition that the regression constants satisfy Assumption~(A1) below---an assumption we tacitly make  throughout this section.}  root-$n$ consistency rates.   OLS estimators under stable errors thus are not even rate-optimal: Proposition~3.1 in Hallin, Swan, Verdebout and Veredas~(2010) indeed establishes  the   local asymptotic normality, with root-$n$ consistency rates, of linear models with stable errors, irrespective of their tail index and skewness parameter.   

This disturbing  fact   
  is by no means   a new finding: see Wise~(1966) or Blattberg and Sargent~(1971) for  early discussion.  Since then, the asymptotic behavior of estimators in linear models with infinite variance and, more specifically, in models with (non Gaussian) stable errors, has attracted much interest, and several alternatives to OLS estimation  have been proposed.  
   Those alternative estimators,  however,  either suffer from major consistency problems, or are strictly inefficient and can be improved: see Section~1.1 for a brief review. The objective of this paper is to show how  {\it one-step R-estimation} allows for a tractable and quite substantial rate-optimal improvement. 
 
 \subsection{Regression parameter estimation under stable errors.}
 Before turning to  R-estimation methods, let us briefly  explain why classical estimation methods fail to provide fully satisfactory solutions. 
 
\begin{enumerate}
\item[(a)] {\it OLS estimators.}  As already mentioned, the main trouble with OLS estimators is that their consistency rate depends on the tail index $\alpha$. This follows from the general results by  Samorodnitsky~\emph{et al.}~(2007) on a class of  linear unbiased estimators   (see point (c) below). That rate is strictly less than the optimal root-$n$ rate, which is a severe drawback. Moreover, the related asymptotic confidence regions  and Wald tests cannot be constructed without estimating $\alpha$ itself. 

\item[(b)] {\it Stable MLEs.} OLS estimators are the maximum likelihood estimators (MLEs) associated with Gaussian likelihoods; better performances can be expected from stable likelihoods (involving  the four parameters of stable densities along with the regression coefficients of interest). A pioneering result by DuMouchel~(1973), indeed,  shows that, somewhat surprisingly, stable MLEs (for location, scale, the tail index  $\alpha$, and the skewness parameter $b$) yield a very standard asymptotically normal  behavior, with traditional root-$n$ rates. This result easily extends to the   regression case\footnote{The situation is quite different for autoregressive and ARMA models   (local experiments are no longer of the LAN type), with  $n^{1/\alpha}$ consistency rates  under tail index $\alpha$,    and  convergence  in distribution  to the maximizer of a random function; see Andrews \emph{et al.}~(2009) for recent results in that context.}. Practical implementation, of course, runs    into the  problem that  non Gaussian stable densities, hence stable likelihoods, cannot be expressed in closed form. For specified tail index $\alpha$ and skewness parameter $b$,  this is not an obstacle anymore thanks to the computationally efficient integral approximations obtained by  Zolotarev~(1986,~1995),  Nolan~(1997,~1999) and several others. But in practice, the tail index and the skewness parameter also have to be estimated; the information matrix, moreover, is not block-diagonal (see DuMouchel~(1975)), so that the estimators $\hat\alpha$ and   $\hat{b}$ of $\alpha$ and  $b$ cannot simply be plugged into the information matrix when confidence regions or Wald tests are to be constructed for the regression parameters. Although asymptotically optimal, stable-likelihood-based inference in practice  thus seems difficult.

\item[(c)] {\it Linear unbiased estimators.} A broad class of linear unbiased estimators, of which OLS estimators are a particular case,  has been considered by Samorodnitsky~\emph{et al.}~(2007), who also provide a quite complete and systematic picture\footnote{Under very general assumptions on the asymptotic behavior of the regression constants (more general than Assumptions~(A1) and~(A2) below), but assuming {\it symmetric} heavy-tailed errors---an assumption we do not make here.} of their asymptotic behavior. Consistency rates, as a rule, crucially depend on the tail index $\alpha$ of the underlying noise, and are strictly less than the optimal root-$n$ ones;  asymptotic covariances depend on $\alpha$ as well. All the drawbacks of OLS estimation thus also are present here. The BLU$\alpha$N   (best linear unbiased estimator, relative to some adequate $\alpha$-norm---limited to $1<\alpha < 2$)   estimators  considered in   El Barmi and Nelson~(1997) suffers from the same problems.

\item[(d)] {\it LAD estimators.} The bad performances of   L$_2$   estimators (OLS) considerably reinforce the attractiveness of  the  L$_1$ approach. The so-called LAD  (Least Absolute Deviations)   estimators (a particular case of more general quantile regression estimators in the Bassett and Koenker~(1978) style) indeed, irrespective of the tail index~$\alpha$, achieve (under Assumption~(A1))   root-$n$ consistency. The asymptotic properties of LAD estimators in regression models have been studied intensively: see  Bassett and Koenker~(1978)   for the standard case, Knight~(1998) or El Bantli and Hallin~(1999) for more general results. Contrary to  stable MLEs, BLUEs and OLS estimators, the LAD ones, thus,  achieve rate-optimal consistency.  Constructing the related confidence regions and Wald tests is possible via classical techniques, without any estimation of $\alpha$. These advantages of LAD estimation in the stable context were emphasized as early as 1971 by Fama and Roll~(1971). On the other hand, LAD estimators, which are optimal under light-tailed double-exponential noise,  cannot be efficient under any heavy-tailed stable density. The objective of this paper is to show how LAD estimators can be improved, often quite substantially, without specifying or estimating  the tail index $\alpha$.   
 
\end{enumerate}

 \subsection{R-estimation under stable errors.}
Estimation methods based on ranks---in short, R-estimation---go back
to Hodges and Lehmann (1963), who provide R-estimators for one-sample
and two-sample location models (under symmetric distributions, for the
one-sample case), based on the Wilcoxon and van der Waerden (signed)
rank statistic. Since then, the technique has been used in a variety
of problems, including $K$-sample location,  regression and analysis
of variance,  time series analysis and elliptical families---see,
e.g., Lehmann~(1963),   Sen~(1966),  Jure\v ckov\' a~(1971), Koul
(1971), Jure\v ckov\' a and Sen~(1996), Koul and Saleh~(1993),  Allal
\emph{et al.} (2001), Koul~(2002), Hallin \emph{et al.}~(2006), Hallin
and Paindaveine~(2008), and many others.  

Ranks naturally appear as maximal invariants in semiparametric models
where the density $f$ of some unobservable noise constitutes the
infinite-dimensional nuisance. Under classical Argmin form, the
Hodges-Lehmann or R-estimator~$\!
\ut\varthetab\phantom{i}\!\!\!\!\!\n_{ 
\text{ HL} } $ of a parameter $\varthetab$ is defined as  
\begin{equation}\label{R-est}  \ut\varthetab\phantom{i}\!\!\!\!\!\n_{
\text{ HL} } := \text{argmin}_{{\bf t}\in\mathbb{R}^K} \left\vert \utQ\!\n ({\bf R}\n({{\bf t}}))\right\vert ,
\end{equation}
where  $\! \utQ\!\n ({\bf R}\n({\varthetab_0}))$ is a (signed)-rank test statistic for the null hypothesis ${\cal H}_0 : \varthetab = \varthetab_0$   (two-sided test). The main advantage of $\!\! \ut\varthetab\phantom{i}\!\!\!\!\!\n_{\text{ HL} } \vspace{-2mm}$ over more usual M-estimators follows from the fact  that (under parameter value $\varthetab$ and error density $f$, and standard root-$n$ consistency conditions),  $n^{1/2}( \ut\varthetab\phantom{i}\!\!\!\!\!\n_{
\text{ HL} } - 
\varthetab) \vspace{-2mm}$ is asymptotically equivalent to a function which depends on the unknown actual density $f$ but is measurable  with respect to the ranks ${\bf R}\n({\varthetab})$ of the unobservable noise (see Hallin and Paindaveine~2008 for details). The asymptotic relative efficiencies (AREs) of the R-estimator~$\! \ut\varthetab\phantom{i}\!\!\!\!\!\n_{
\text{ HL}}\vspace{-2mm}$ defined in (\ref{R-est}) with respect to  other R-estimators, or with respect to its Gaussian competitor (OLS or Gaussian MLE, whenever the latter are root-$n$ consistent) are the same as the AREs of the corresponding rank tests with respect to their Gaussian competitors.\footnote{ Since Gaussian methods are generally invalid under stable error densities, AREs in the sequel are taken with respect to double-exponential likelihood procedures, that is, least absolute deviation (LAD) estimators and the regression version of sign tests (the Laplace rank tests).} 

The Argmin form (\ref{R-est}), however, is computationally
inconvenient--particularly so in the case of a relatively
high-dimensional parameter $\varthetab$. Inspired by Le Cam's one-step
estimation method, Hallin {\it et al.}~(2006), in the context of
R-estimation of   shape matrices in elliptical families and Hallin and
Paindaveine~(2008, unpublished manuscript), in a more general context, therefore
introduced a one-step form of R-estimation. That method, contrary to
(\ref{R-est}), avoids the computational inconvenience of minimizing,
over a possibly high-dimensional parameter space,   a piecewise
constant function of the form $\vert \utQ\!\n ({\bf R}\n({{\bf
    t}}))\vert$;  moreover it also provides, as a by-product, the
asymptotic covariance matrix of the R-estimator.  On the other hand,
one-step methods require the existence of a preliminary rate-optimal
consistent (here,  root-$n$ consistent) estimator. This role will be
played, in the present context, by the LAD estimator, the only one in
the existing literature enjoying  the required consistency
properties. Our R-estimators thus  appear  as a one-step improvements
over the LAD estimators; they yield the same collection of ARE values
as the corresponding rank-based tests, the values of which were
obtained in Hallin {\it et al.}~(2010).

 In this paper, we explain how that one-step method can be implemented for the estimation of the regression parameter of a general linear model with stable errors, and we study the asymptotic performances of the resulting  R-estimators . Those R-estimators rely on a rank-based version of Le Cam's one-step methodology which bypasses the nonparametric estimation of cross-information quantities. 
They are asymptotically normal under any stable density (with standard root-$n$ rate), and efficient at some prespecified stable density~$f_{\thetab}$. They exhibit the same asymptotic relative efficiencies as the rank-based tests studied in Hallin \emph{et al.} (2010).  For specific scores, they outperform   LAD estimators,  and hence   all valid and tractable estimation methods proposed in the literature. In particular, when based on certain stable scores, such as  the score associated with the symmetric stable distribution with tail parameter $\alpha=1.4$ (see Figure \ref{figAREs2}), they dominate the LAD under any stable distribution with $\alpha\in (1,2)$.  The computational advantages  of   one-step R-estimators  over the more classical Argmin ones   lie in the fact that  the $K$-dimensional minimization (\ref{R-est}) of a non convex piecewise constant rank-based objective function is replaced by the minimization of a continuous, strictly convex L$_1$ criterion (yielding the preliminary LAD estimator), followed by a one-dimensional optimization problem; the LAD estimator, moreover, can be obtained exactly as the solution of a linear programming problem. Table~\ref{Tab3} below provides numerical evidence of the quite substantial advantages (in terms of bias and mean squared error) of one-step R-estimation over its    classical Argmin counterpart.  

%

\section{R-estimation of regression coefficients.}

\subsection{Asymptotics for  linear  models with stable errors.}

 The family of $\alpha$-stable densities is a four-parameter family 
 $$\Big\{f_{\thetab} = f_{\alpha, b, \gamma, \delta} \vert \ \thetab:=(\alpha, b, \gamma, \delta)\pr\in\Thetab = (0, 2] \times [-1,1]\times \R^+\!\! \times \R\Big\}.$$ 
 Writing $f_{\alpha, b}$   for    $f_{\alpha, b, 1, 0}$, we   have  
\begin{equation}\label{locationscale}f_{\alpha, b,  \gamma, \delta}(x) =\frac{1}{\gamma} f_{\alpha, b}\left(\frac{x-\delta}{ \gamma}\right) ,\end{equation}
which characterizes the roles of $\delta$ and $\gamma$ as location and scale parameters, respectively, and that of  $f_{\alpha, b}$  as  the    standardized version of  $ f_{\alpha, b, \gamma, \delta}$. The   parameters $\alpha$ and $b$ determine the shape of the distribution, with $\alpha$ being the \emph{characteristic exponent} (or \emph{tail index}) and $b$  the \emph{skewness} parameter---an interpretation  justified by  the   fact that,  for $b=0$, $f_{\alpha, b, \gamma, \delta}$ is symmetric  with respect to $\delta$ and, for $0<b\leq1$ (resp., $-1\leq b<0$),   skewed to the right  (resp., to the left)--- see Section~1.2 of Samorodnitsky and Taqqu~(1994) for  details.    The notations~$F_{\thetab}$ and  $F_{\alpha, b, \gamma, \delta}$ will be used for the  distribution function associated with $f_{\thetab}$. 

Some particular choices of $\thetab$ yield well-known distributions, namely   the Gaussian ($\alpha=2$, any $b$), the Cauchy ($\alpha=1$, $b=0$) and the L\'evy ($\alpha=1/2$, $b=1$). However, together with the reflected L\'evy density, these are  the only instances of stable densities  that can be expressed  explicitly in terms of elementary functions. For all other  choices of the parameters,   a closed form for $f_\thetab$ is not possible, and stable distributions  either are defined   in terms of       characteristic functions and inverse Fourier transforms, or via integral formulas (see e.g. Nolan~(1997) or Zolotarev~(1986)).

Throughout, we  consider a vector     ${\bf X}\n:=(X_1^{(n)}, \ldots, X_n^{(n)})\pr$ of observations satisfying  
\begin{equation} \label{model} X_{i}^{(n)} = a+ \sum_{k=1}^Kc_{ik}^{(n)}\beta_k+\epsilon_i^{(n)},\, \, i = 1, \ldots, n, \end{equation}
for some intercept $a\in\R$ and   the regression parameters $\betab :=(\beta_1, \ldots, \beta_K)\pr\in\R^K$;   $c_{i1}^{(n)}, \ldots, c_{iK}^{(n)}$ ($i=1, \ldots,n$) are   regression constants,   and $\{ \epsilon_i^{(n)} , i\in\N\} $  is a sequence of nonobservable  i.i.d. errors  with stable  density~$f_{\thetab}$,~$\thetab = (\alpha, b,  \gamma, 0)  \in  \Thetab$. 

The construction of our R-estimators is based on the uniform local asymptotic normality (ULAN) property, with respect to $\betab$, of the regression model (\ref{model}) under stable error densities. That property is  established in  Hallin \emph{et al.}~(2010) under the following technical assumptions.  Without loss of generality, we  impose that  $\sum_{i=1}^nc_{ik}^{(n)}=0$ for $k=1, \ldots, K$; letting ${\bf c}\n_i:=(c\n_{i1},\ldots , c\n_{iK})\pr$, $ \mathbb{C}^{(n)} := n^{-1}\sum_{i=1}^n{\bf c}_i^{(n)}{{\bf c}_i^{(n)}}\pr$,   we make the following assumptions on the asymptotic behavior of the regression constants.  \vspace{2mm}

\noindent{\sc Assumption (A1)} For all $n\in \N$, $\mathbb{C}^{(n)}$ is positive definite and converges, as $n\to\infty$, to a positive definite matrix~$\mathbb{K}^{-2}$. 
\vspace{2mm}

\noindent {\sc Assumption (A2)} (Noether conditions) For all $k=1, \ldots, K$, one has 
$$\lim_{n\to\infty} \Big\{{\displaystyle{\max_{1\le t \le n} \left(c_{tk}^{(n)}
 \right)^2}} {\Big /} \ {\displaystyle{ \sum_{t=1}^n\left(c_{tk}^{(n)}
 \right)^2}} \Big\}= 0.\vspace{2mm}
$$

Denoting by ${\rm P}\n_{\thetab , a, \betab }$ the probability distribution of ${\bf X}\n$ under (\ref{model}),  let 
$$ Z_i^{(n)}(\betab) := X_i^{(n)} - a- \sum_{k=1}^Kc_{ik}^{(n)}\beta_k ,\qquad i=1,\ldots , n$$ 
stand for  the residuals associated with the value $\betab$ of the regression parameter: under ${\rm P}\n_{\thetab , a, \betab }$, the $ Z_i^{(n)}(\betab)$'s thus are i.i.d. with density $f_{  (\alpha, b,  \gamma, 0) }$. 
 Here and in the sequel, we write $ Z_i^{(n)}(\betab)$ instead of $ Z_i^{(n)}(a, \betab)$  for the sake of simplicity. Although the quantity appearing in Proposition~\ref{reglan}   depends on $a$,  the rank-based statistics $\utDelta_{J}^{(n)}$ defined in~(\ref{rankbased}) below do not, as   the  $ Z_i^{(n)}(\betab)$'s  only enter the definition through their ranks, which do not depend on $a$ (fortunately so, as $a$  remains an unspecified nuisance).   The  following result is proved  in Hallin \emph{et al.} (2010).

\begin{proposition}\label{reglan} {\rm (ULAN,  Hallin, Swan, Verdebout and Veredas
  2010).} Suppose that Assumptions (A1) and~(A2) hold. Fix $\thetab = (\alpha, b, \gamma  , 0) \in \Thetab$.   
Then, model (\ref{model}) (the family $\{{\rm P}\n_{\thetab , a, \betab }\vert \ \betab\in\R^K\}$),  is ULAN with respect to   $\betab$, with contiguity rate $n^{1/2}$. More precisely,  letting ${\pmb \nu}(n):= n^{-\frac{1}{2}}  \mathbb{K}^{(n)}$ with $\mathbb{K}^{(n)} := \left(\mathbb{C}^{(n)}\right)^{-1/2}$, for all $\betab\, \in \, \R^K$, all sequences~$\betab\n$ such that ${\pmb \nu}^{-1}(n)(\betab\n - \betab ) = O(1)$  and all bounded sequences ${\pmb \tau}^{(n)} \in \R^{K}$, 
$$\begin{array}{rcl}
\Lambda^{(n)}_{\thetab, \betab\n +{\pmb \nu}(n)\taub^{(n)}}  & := &\log\left[
{\DS{{\rm d}{\rm P}\n_{\thetab , a, \betab\n+{\pmb \nu}(n)\taub^{(n)} }}}
/
{\DS{{\rm d}{\rm P}\n_{\thetab , a, \betab\n }}} \right]\vspace{3mm}
\\
&=& \log\left[
{\prod_{t=1}^n\DS{ f_{\thetab}( Z_t^{(n)}( \betab+{\pmb \nu}(n)\taub^{(n)}))}}
/
{\prod_{t=1}^n \DS{ f_{\thetab}( Z_t^{(n)}(\betab))}}\right] \\
											 & & \\
											  &  = & {{\taub}^{(n)}}\pr  {{\Deltab}^{(n)}_{\thetab}}(\betab\n) -\frac{1}{2} {{\taub}^{(n)}}\pr  {\taub}^{(n)}\mathcal{I}(\thetab) +o_P(1)
											    \end{array}$$
under $\mathcal{H}_{\thetab}^{(n)}(\betab)$	as $n\to\infty$, where, setting $\varphi_{\thetab}  := - \dot{f}_{\thetab} /f_{\thetab} $, with $\dot{f_\thetab}$ the derivative of $x\mapsto{f_\thetab(x)}$ and 
$$\mathcal{I}(\thetab):= \int_{-\infty}^\infty \varphi^2_{\thetab}(x)f_{\thetab}(x){\rm d}x,$$
$\mathcal{I}(\thetab){\bf I}_K$ is the {information matrix} and 
\begin{equation}\label{centralseq} {{\bf \Delta}^{(n)}_{\thetab}}(\betab) := n^{-1/2}\mathbb{K}^{(n)\prime}\sum_{i=1}^n \varphi_{\thetab}\left(Z_i^{(n)}(\betab)\right){\bf c}_i^{(n)}\stackrel{\mathcal{L}}{\longrightarrow} \mathcal{N}\big({\bf 0},\  \mathcal{I}(\thetab){\bf I}_K\big) \end{equation}
the {central sequence}.\end{proposition}

  ULAN,  here as  in Hallin {\emph{et al.}} (2010),  is stated  under stable distributions, but of course is well known to hold under any density $f$ such that $f^{1/2}$ is differentiable in quadratic mean;  ${\rm P}\n_{\thetab , a, \betab }$, $\varphi_{\thetab} $ and $\mathcal{I}(\thetab)$  then are to be replaced with ${\rm P}\n_{f ,  a,\betab }$, $ \varphi_{f} := 2D(f^{1/2})/f^{1/2}$ and $\mathcal{I}_f$, where $D(f^{1/2})$ stands for the quadratic mean derivative of $f^{1/2}$ and $\mathcal{I}_f:=\int _{-\infty}^\infty  \varphi_{f}^2(x)f(x){\rm d}x$.  Denote by $\cal F$ that class of densities and  by $\Deltab_{f}\n(\betab)$ the corresponding central sequences. 
\subsection{One step R-estimators.}

The vector ${\bf R}^{(n)} = {\bf R}^{(n)}(\betab) := (R_1^{(n)}, \ldots, R_n^{(n)})$, where  $R_i^{(n)}=R_i^{(n)}(\betab )$ denotes the rank of the residual $Z_i^{(n)}=Z_i^{(n)}(\betab) $, $i=1, \ldots,n$, among $Z_1^{(n)}, \ldots, Z_n^{(n)}$, is distribution-free as  $f $ and $a$ range over the class of all nonvanishing densities and $\R$, respectively.   Throughout, we consider the class of  rank-based   statistics 
\begin{equation}\label{rankbased}\utDelta_{J}^{(n)}(\betab) := n^{-\frac{1}{2}}\mathbb{K}^{(n)\prime}\sum_{i=1}^nJ\left(\frac{R_i^{(n)}}{n+1}\right){\bf c}_i^{(n)},\end{equation}
where $J:\,(0,1)\to\R$ is some score generating function satisfying  \vspace{.2cm}

\noindent {\sc Assumption (B)} The score function $J:\,(0,1)\to\R$ is not constant, and the difference $J_1-J_2$ between two right-continuous and square integrable non-decreasing monotone functions $J_{1}$ and $J_2 : \, (0,1) \to \R$. \vspace{3mm}

 Strongly unimodal densities  $f$ trivially  satisfy that assumption.\footnote{A density $f$ is called  {\it strongly unimodal} if  $f^{1/2}$ is differentiable in quadratic mean and $\varphi_f$ is monotone increasing; Gaussian, logistic and double exponential densities are strongly unimodal. }  Except for the Gaussian one, stable densities (\ref{centralseq})  are not strongly unimodal. However, $u\mapsto  \varphi_f(F^{-1}(u))$ being   bounded (in absolute value) and continuously differentiable,  with a derivative changing signs exactly twice, it has bounded variation, hence can be expressed as the difference between two monotone increasing functions; $\varphi_f$ therefore also can.    \vspace{.2cm}

The following result summarizes the asymptotic properties of the rank-based statistics (\ref{rankbased}); see the Appendix for a proof. 

\begin{proposition} \label{asymplin}Let Assumptions  (A1), (A2) and (B) hold. Then, 
\begin{itemize}
\item[(i)] letting ${\pmb{\Delta}}_{J }^{(n)}(\betab) := n^{-\frac{1}{2}}\mathbb{K}^{(n)\prime}\sum_{i=1}^nJ(G  (Z_i^{(n)}(\betab) ) ){\bf c}_i^{(n)}, $ where $G$ stands for the distribution function  associated with  a density $g \in \mathcal{F}$, we have, under $ {\rm P}\n_{g, a , \betab}$, as $\ny\vspace{-2mm}$, 
\begin{equation}\label{Hajek}
\utDelta_{J}^{(n)}(\betab) 
-{\pmb{\Delta}}_{J}^{(n)}(\betab) = o_{\rm P}(1).
\end{equation}
 Hence,  for  $J(u)= \varphi_{f}(F^{-1}(u))$ with $f \in \mathcal{F}$,${{\ut\Delta}\ }{\hspace{-2.7mm}}_{J }^{(n)}(\betab)$  is asymptotically equivalent,\footnote{Since central sequences   are only defined up to  $o_{\rm P}(1)$ terms,${\ut\Delta}\ {\hspace{-2.7mm}}_{J }^{(n)}(\betab)\vspace{-2mm} $ thus is a rank-based version of the central sequence~$\Deltab_{f}\n(\betab)$.} under   $ {\rm P}\n_{f, a , \betab}\vspace{-2mm}$, to~$\Deltab_{f}\n(\betab)$;
\item[(ii)] under $ {\rm P}\n_{g, a , \betab}$ ($g \in \mathcal{F}$), $\utDelta_{J}^{(n)}(\betab) $ is asymptotically normal with mean zero and covariance\vspace{1mm} matrix $\mathcal{J}(J) {\bf I}_K$, where 
$\mathcal{J}(J) :=\int_{0}^1 J^2(u){\rm d}u;$
\item[(iii)] under $ {\rm P}_{g, a , \betab+ \nub\n{\taub}}$ ($g \in \mathcal{F}$),  $\utDelta_{J}^{(n)}(\betab) \vspace{1mm}$ is asymptotically normal with mean   $\mathcal{J}(J, g){\taub}$ and covariance matrix 
$\mathcal{J}(J) {\bf I}_K$,  where 
\begin{equation}\mathcal{J}(J, g) :=\int_{0}^1 J(u)\varphi_{g}(G^{-1}(u)){\rm d}u;\end{equation}\label{calJ}
\item[(iv)] $\utDelta_{J}^{(n)}(\betab) $ satisfies the asymptotic linearity property
\begin{equation} \label{asymplineq} \utDelta_{J}^{(n)}(\betab+\nub\n{\taub}\n)-\utDelta_{J}^{(n)}(\betab) = -\mathcal{J}(J,g) {\taub}\n + o_{\rm P}(1) \end{equation} under $ {\rm P}\n_{g, a , \betab}$ with $g \in \mathcal{F}$, as $\ny$.\vspace{2mm}

\end{itemize}

\end{proposition} 

Under the conditions of Proposition~2.1, the Le Cam one-step methodology requires the existence of a preliminary root-$n$ consistent estimator $\hat{\betab}\!\!\phantom{^I}\n$ of~$\betab$.  The LAD  estimator  $\hat{\betab}\phantom{i}\!\!\n_{\scriptscriptstyle{\text{\rm LAD}}}$ of $\betab$, which we are considering in the sequel,  is one possibility,   but any other estimator enjoying  root-$n$ consistency under the whole class of stable densities  would be an equally valid  candidate. 

The LAD estimator   $(\hat{a}\n_{\scriptscriptstyle{\text{\rm LAD}}}, \hat{\betab}^{(n)\prime}_{\scriptscriptstyle{\text{\rm LAD}}})\pr$ of $(a,\betab\pr)\pr$  is obtained by minimizing the $L_1$-objective function
$$(\hat{a}\n_{\scriptscriptstyle{\text{\rm LAD}}}, \hat{\betab}^{(n)\prime}_{\scriptscriptstyle{\text{\rm LAD}}})\pr:= {\rm argmin}_{(a, \betab) \in \mathbb{R}^{K+1}} \sum_{i=1}^n | Z_i\n(\betab) |.$$
In this context, however, $a$ needs not be estimated, as ranks are insensitive to location shift; we therefore concentrate on $\hat{\betab}^{(n)}_{\scriptscriptstyle{\text{\rm LAD}}}$.  In order to control for the uniformity of local behaviors, a discretized version $\hat{\betab}\n_{\#}$ of~$\hat{\betab}\phantom{i}\!\!\n_{\scriptscriptstyle{\text{\rm LAD}}}$
should be considered in theoretical asymptotic statements.  The discretization trick, which is 
due to Le Cam, is quite standard in the context of one-step estimation. While retaining root-$n$ consistency, discretized estimators indeed enjoy the important property 
of  {\it asymptotic local discreteness}, that is, they only take a finite 
number of distinct values, as~$\ny$,  in $\betab$-centered balls with $O(n^{-1/2} )$ radius. In fixed-$n$ practice, however, such discretizations are irrelevant (the discretization constant can be chosen arbitrarily large). For the sake of simplicity, we will henceforth tacitly assume that $\hat{\betab}\phantom{i}\!\!\n_{\scriptscriptstyle{\text{\rm LAD}}}$, in asymptotic statements,  has been adequately discretized. 

Were  $\mathcal{J}^{-1}(J, g)$ a known quantity, the one-step R-estimator of $\pmb\beta$ would  take (since the asymptotic variance of $\utDelta_{J}^{(n)}$ is proportional to an identity matrix) the following  very simple form: 
  \begin{equation} \label{ourestimate}
\tilde{\utbeta}\!\!\!\phantom{^I}\n_{J}:= \hat{\betab}\phantom{i}\!\!\n_{\scriptscriptstyle{\text{\rm LAD}}}+ \nub\n\mathcal{J}^{-1}(J, g) \utDelta_{J}^{(n)}(\hat{\betab}\phantom{i}\!\!\n_{\scriptscriptstyle{\text{\rm LAD}}}). \end{equation}
It readily follows from (\ref{asymplineq})  (as well as from standard results on one-step estimation: see, e.g., Proposition~1 in Chapter~6 of Le~Cam and Yang~(2000)) that 
$${\pmb\nu}^{-1}(n)(\tilde{\utbeta}\!\!\!\phantom{^I}\n_{J} - {\betab}) = \mathcal{J}^{-1}(J, g)\utDelta_{J}^{(n)}(\betab )+o_{\rm P}(1),$$
hence, that ${\pmb\nu}^{-1}(n)(\tilde{\utbeta}\!\!\!\phantom{^I}\n_{J} - {\betab}) $ is asymptotically ${\cal N}({\bf 0}, ( \mathcal{J}(J)/  \mathcal{J}^{2}(J, g)){\bf I}_K)$ under $ {\rm P}\n_{g, a , \betab}$ ($g \in \mathcal{F}$). This  in turn implies that ${\pmb\nu}^{-1}(n)(\tilde{\utbeta}\!\!\!\phantom{^I}\n_{J} - {\betab}) $,  for  $J(u)= \varphi_{f}(F^{-1}(u))$,  is asymptotically ${\cal N}({\bf 0},    \mathcal{J}^{-1}(J){\bf I}_K)$ under $ {\rm P}\n_{f, a , \betab}$, that is, reaches parametric efficiency at correctly specified density $f=g$. \vspace{2mm}

Unfortunately,  the scalar {\it cross-information quantity} $\mathcal{J}(J, g)$ is not known---a phenomenon that does not appear in the usual one-step method, based on the  ``parametric central sequence" associated with some correctly identified density $f=g$. Under definition (\ref{ourestimate}),  $\tilde{\utbeta}\!\!\!\phantom{^I}\n_{J}$ therefore  is not a genuine estimator. That cross-information quantity $\mathcal{J}(J, g)$ thus has to be consistently estimated. To obtain such a consistent estimator, we adopt here the idea first developed in Hallin  \emph{et al.} (2006) and generalized in Cassart \emph{et~al.}~(2010). \vspace{2mm}

For all $v>0$, define $\tilde{\betab}\!\!\phantom{^I}\n(v):= \hat{\betab}\phantom{i}\!\!\n_{\scriptscriptstyle{\text{\rm LAD}}}+ \nub\n v  \utDelta_{J}^{(n)}(\hat{\betab}\phantom{i}\!\!\n_{\scriptscriptstyle{\text{\rm LAD}}})$,
and consider the scalar product 
$$
 h\n(v):= 
  \big(\utDelta_{J}^{(n)}(\hat{\betab}\phantom{i}\!\!\n_{\scriptscriptstyle{\text{\rm LAD}}}) \big)^{\prime}  \utDelta_{J}^{(n)}(\tilde{\betab}\n(v)).
$$
Proposition \ref{asymplin}, the consistency and local asymptotic discreteness of $\hat{\betab}\phantom{i}\!\!\n_{\scriptscriptstyle{\text{\rm LAD}}}$, and the definition of $\tilde{\betab}\!\!\phantom{^I}\n(v)$
entail that, under $ {\rm P}\n_{g, a , \betab}$ with~$g \in \mathcal{F}$,  
\begin{eqnarray}\nonumber
   h\n(v) 
  & = & 
 \left(  \utDelta_{J}^{(n)}(\hat{\betab}\phantom{i}\!\!\n_{\scriptscriptstyle{\text{\rm LAD}}})\right)\pr \left(  \utDelta_{J}^{(n)}(\hat{\betab}\phantom{i}\!\!\n_{\scriptscriptstyle{\text{\rm LAD}}}) - \mathcal{J}(J, g) \; n^{1/2} ({\mathbb K}^{(n)})^{-1} (\tilde{\betab}\n(v)- \hat{\betab}\phantom{i}\!\!\n_{\scriptscriptstyle{\text{\rm LAD}}})\right) +o_{\rm P}(1)\\ 
 &=& \left(  \utDelta_{J}^{(n)}(\hat{\betab}\phantom{i}\!\!\n_{\scriptscriptstyle{\text{\rm LAD}}})\right)\pr \left(  \utDelta_{J}^{(n)}(\hat{\betab}\phantom{i}\!\!\n_{\scriptscriptstyle{\text{\rm LAD}}}) -  \mathcal{J}(J, g) v\utDelta_{J}^{(n)}(\hat{\betab}\phantom{i}\!\!\n_{\scriptscriptstyle{\text{\rm LAD}}})\right) +o_{\rm P}(1)\nonumber 
 \\ 
   &=& 
   (1- \mathcal{J}(J, g) v)  \; h\n(0) +o_{\rm P}(1)\label{hfun}
\end{eqnarray}
 for any $v>0$;   this provides the intuition for taking the solution of $h(v) =0$ as an estimation  of $(\mathcal{J}(J, g))^{-1}$.    And, provided that $h\n(0)$ is not $o_{\rm P}(1)$, a consistent estimator of $(\mathcal{J}(J, g))^{-1}$   indeed would be 
$$
\hat{v}\n := {\rm inf} \{ v >0 : h\n(v) <0 \}.
$$

More precisely, consider a discretization of the positive half-line, with $v_\ell :=\ell /c$, $\ell\in \N$, $c>0$ a (typically, large) discretizing constant,  the value of which, however,  plays no role in asymptotic statements. Putting 
\begin{equation}\label{hatvpm}
v\n_-:=\text{min}\{\ell \text{ such that } h\n (v\n_{\ell + 1}) < 0
\} \quad \text{ and }\quad v\n_+:=v\n_- + \frac{1}{c},
\end{equation}
consider the linear interpolation
\begin{equation}\label{hatvn}
\hat{v}\n:=
v\n_-\left(1 - \frac{h\n(v\n_-)}{h\n(v\n_-) - h\n(v\n_+)}
\right)
+
v\n_+\frac{h\n(v\n_-)}{h\n(v\n_-) - h\n(v\n_+)}.
\end{equation}
It follows from  Proposition~2.1 in Cassart \emph{et al.}~(2010)   that, unless $h\n(0)$ is  $o_{\rm P}(1)$,   $\widehat{\mathcal{J}}(J, g):=(\hat{v}\n)^{-1}$ provides a consistent estimator of the cross-information quantity ${\mathcal{J}}(J, g)$. 
    Our one-step R-estimator then is defined as\vspace{1mm}
  $${\utbeta}\!\!\!\phantom{^I}\n_{J}:=\tilde{\betab}\!\!\phantom{^I}\n(\widehat{\mathcal{J}}^{-1}(J, g)) = \hat{\betab}\phantom{i}\!\!\n_{\scriptscriptstyle{\text{\rm LAD}}}+ \nub\n \widehat{\mathcal{J}}^{-1}(J, g)\utDelta_{J}^{(n)}(\hat{\betab}\phantom{i}\!\!\n_{\scriptscriptstyle{\text{\rm LAD}}})
 . $$
 Now, if $J$ is such that $\utDelta_{J}^{(n)}(\hat{\betab}\phantom{i}\!\!\n_{\scriptscriptstyle{\text{\rm LAD}}}) =o_{\rm P}(1)$, that is, if the Laplace or double-exponential score  function $u\mapsto J_{\text{L}}(u):=\sqrt2 \, {\rm sign}(u-1/2)$,  is considered,  we have (see Proposition~2.4)  ${\utbeta}\!\!\!\phantom{^I}\n_{J_L}=  \hat{\betab}\phantom{i}\!\!\n_{\scriptscriptstyle{\text{\rm LAD}}}+o_{\rm P}(n^{-1/2})$ and $\tilde{\utbeta}\!\!\!\phantom{^I}\n_{J_\text{L}}= \hat{\betab}\phantom{i}\!\!\n_{\scriptscriptstyle{\text{\rm LAD}}}+o_{\rm P}(n^{-1/2})$,  so that our estimator coincides, asymptotically, with the LAD estimator.  
 
 The following result (see the Appendix for a proof) summarizes the asymptotic properties of ${\utbeta}\!\!\!\phantom{^I}\n_{J}$.
 
 \begin{proposition} \label{asympnorm}
 
 Let Assumptions  (A1), (A2) and (B)   hold.  Then, 
 $n^{1/2}({\utbeta}\!\!\!\phantom{^I}\n_{J}\! - \betab)$ is asymptotically normal with mean zero and covariance matrix $ \big({\mathcal{J}(J)}/{\mathcal{J}^2(J, g)} \big)\mathbb{K}^2$ under $ {\rm P}\n_{g, a , \betab}$ with $g \in \mathcal{F}$.  Therefore, letting   $J(u)= \varphi_{f}(F^{-1}(u))$, ${\utbeta}\!\!\!\phantom{^I}\n_{J}$ achieves the parametric  efficiency bound under $ {\rm P}\n_{f, a , \betab}$.
 \end{proposition}

In view of Proposition \ref{asympnorm},  the asymptotic relative efficiencies of our R-estimators   clearly coincide with those of the corresponding tests developed in Hallin \emph{et al.}~(2010). 
More precisely,  we have that   
\begin{equation}\label{AREf}  \mbox{ARE}_{g}(J_1/J_2)  = {\mathcal{J}^2(J_1, g)}
 {\mathcal{J}(J_2)} /
 {\mathcal{J}^2(J_2, g)}{\mathcal{J}(J_1)},
 \end{equation}
where {ARE}$_{g}(J_1/J_2) \vspace{-1mm}$ denotes the asymptotic relative efficiency, under  density $g$, of the R-estimator ${\utbeta}\!\!\!\phantom{^I}\n_{J_1}$, based on the score-generating function $J_1$, with respect to the R-estimator ${\utbeta}\!\!\!\phantom{^I}\n_{J_2}\vspace{-1mm}$, based on the score-generating function $J_2$.

 \begin{table}[htbp]
\begin{center}
\caption{\normalsize AREs of R-estimators with respect to LAD estimators} \vspace{0.15cm}
{\small
\begin{tabular}{ccccc}
\hline
\hline
\vspace{0.1cm} Estimators & \multicolumn{4}{c}{Underlying stable density}\\
\hline
 & $\alpha=2$; $b=0$ & $\alpha=1.8$; $b=0$& $\alpha=1.8$; $b=0.5$  & $\alpha=0.5$; $b=0.5$ \vspace{0.1cm}\\ 
${\utbeta}\n_{J_{\rm W}}/\hat{\betab}\n_{\text{\rm LAD}}$   &  1.4999    & 1.3888     & 1.3984        &  1.7776 \\ 
${\utbeta}\n_{J_{\rm vdW}}/\hat{\betab}\n_{\text{\rm LAD}}$ &  1.5708    & 1.3056     &  1.3285         &  1.251 \\
${\utbeta}\n_{J_{\rm C}}/\hat{\betab}\n_{\text{\rm LAD}}$ &  0.6759     & 0.7880     &  0.7769        & 2.007 \\
${\utbeta}\n_{J_{\rm 1.8; 0}}/\hat{\betab}\n_{\text{\rm LAD}}$ &1.4459  & 1.4183 & 1.4222  & 1.6453 \\
${\utbeta}\n_{J_{\rm 1.8; .5}}/\hat{\betab}\n_{\text{\rm LAD}}$& 1.4452 & 1.3969  &  1.4459  &1.4432  \\
${\utbeta}\n_{J_{\rm .5 ; .5}}/\hat{\betab}\n_{\text{\rm LAD}}$ &  0.0925& 0.1099  &0.1175    & 21.2364\vspace{2mm} \\
\hline
\hline
\multicolumn{5}{p{12cm}}{AREs for R-estimators based on various scores with respect to the LAD estimator.  Columns  correspond to  the (stable) densities under which AREs are computed, rows   to the  scores considered: Wilcoxon (${J_{ \protect \rm W}}$), van der Waerden (${J_{ \protect \rm vdW}}$), Cauchy (${J_{ \protect \rm C}}$), and three ($\delta=0, \; \gamma=1$) stable scores (${J_{ \protect \alpha ;  b}}$); recall that the R-estimator based on Laplace scores asymptotically coincides with the LAD estimator (see Proposition \ref{LADvsLap}).\label{tableAREs2}}\
\end{tabular}
}
\end{center} 
\end{table}

\begin{figure}[htbp]
\begin{center}
\includegraphics[scale=.55]{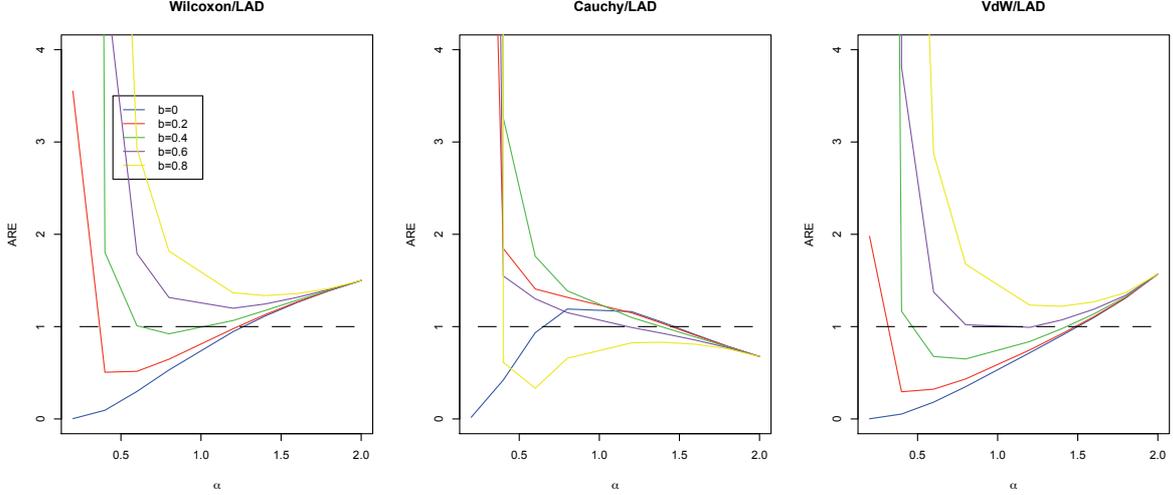}
\end{center}\vspace{-8mm}
\caption{\normalsize AREs of R-estimators based on  Wilcoxon, Cauchy and van der Waerden scores, with respect to the LAD estimator, as a function of $\alpha$ and for various values of $b$.\label{figAREs1}}
\end{figure}

\begin{figure}
\begin{center}
\includegraphics[scale=.55, angle=90]{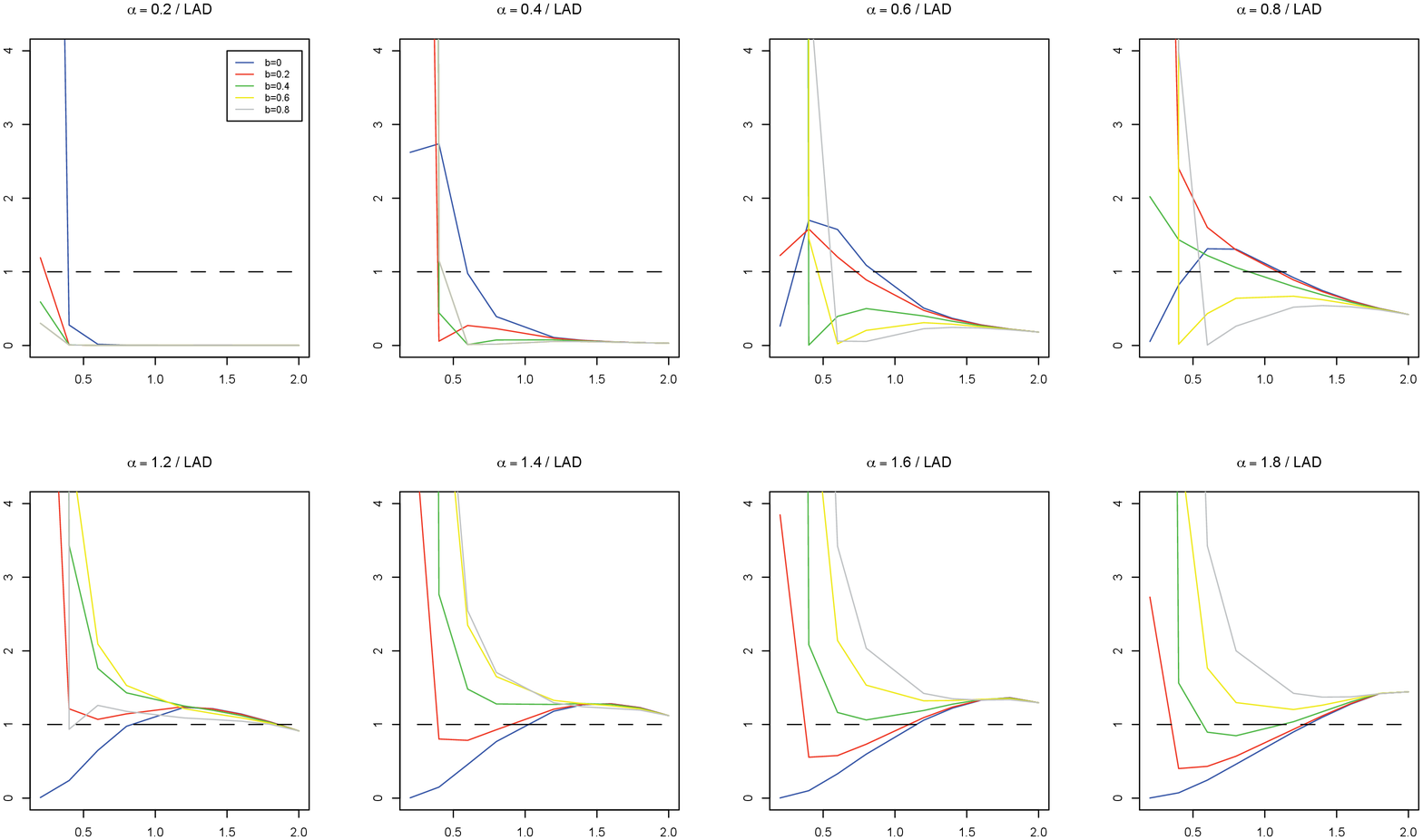}
\end{center}
\vspace{-4mm}

\caption{\normalsize AREs under stable distributions of R-estimators based on various stable scores with respect to the LAD estimator, as a functions of $\alpha$ and $b$. \label{figAREs2}}
\end{figure}

 Traditional scores (such as the van der Waerden,   Wilcoxon and   Laplace ones) are associated with some classical light-tailed densities (such as the normal, logistic and double-exponential), leading to   the score-generating functions   
\begin{equation*} 
 J_{\text{vdW}}(u)= \Phi^{-1}(u), \quad J_{\text{W}}(u) = \frac{\pi}{\sqrt{3}}(2u-1), \quad \mbox{ and } \quad J_{\text{L}}(u)= \sqrt2 \, {\rm sign}(u -1/2 ),\end{equation*}
respectively, where 
$\Phi$ denotes, as usual, the  standard normal distribution function. The resulting R-esti\-mators are reaching  parametric efficiency  under  Gaussian,  logistic and  double-exponential densities, respectively.   {\it Stable scores}, of the form  $J_\thetab(x) = - \dot{f}_\thetab(F_\thetab^{-1}(x))/ f_\thetab(F_\thetab^{-1}(x))$, where $f_\thetab$ is some stable density, also can be considered, not under closed form, though; we refer to Appendix~B of Hallin \emph{et al.}~(2010), where rank tests based on such stable scores are discussed, for details.  Table \ref{tableAREs2} and Figures~\ref{figAREs1} and \ref{figAREs2} provide numerical values of AREs in (\ref{AREf}) for various estimators and underlying stable densities. Interestingly, the R-estimators based on the stable scores for tail index 1.4 uniformly dominate, irrespective of the asymmetry parameter $b$, the LAD estimator for all values of $\alpha\in [1,2]$. Their AREs with respect to LAD estimators moreover culminates  in the vicinity of $\alpha=1.8$, a value which is generally recognized as a reasonable tail index for financial data.\footnote{Dominicy and Veredas (2010) found that the estimated $\alpha$ for 22 major worldwide market indexes (nine years of daily returns) ranges between 1.55 to 1.90, with an average of 1.75. Similar values have been  obtained for other financial assets, e.g.    in  Mittnik \emph{et al.}~(2000) or  Deo (2002).}


To conclude this section, the following result establishes the asymptotic equivalence between the LAD estimator and the Laplace R-estimator (based on the score function $J_{\text{L}}$); see the Appendix for a proof.
\begin{proposition} \label{LADvsLap} Let Assumptions (A1) and (A2) hold. Then, the difference ${\utbeta}\!\!\!\phantom{^I}\n_{{J}_{\rm{L}}} - \hat{\betab}\phantom{i}\!\!\n_{\scriptscriptstyle{\text{\rm LAD}}}$ is $o_{\rm P}(n^{-1/2})$ as~$\ny$ under  ${\rm P}\n_{g, a , \betab}$ for any ${g \in \mathcal{F}}$ such that  $g$ is strictly positive at the median $G^{-1}(\frac{1}{2})$.
\end{proposition}

As a direct consequence, the ARE (under ${\rm P}\n_{g, a , \betab}$ with ${g \in \mathcal{F}}$) of any estimator $\tilde{\betab}\n$ with respect to $ \hat{\betab}\phantom{i}\!\!\n_{\scriptscriptstyle{\text{\rm LAD}}}$ is equal to the ARE of 
 $\tilde{\betab}\n$ with respect to ${\utbeta}\!\!\!\phantom{^I}\n_{{J}_{\rm{L}}}$.

\section{Finite-sample performance.}

 This  section is devoted to a simulation study of the finite-sample performances of the various R-estimators described in the previous sections and some of their competitors, in order to check whether  these performances are are in line with the ARE results of Table~1. 

 We generated $M=1000$ samples  from two multiple  regression models, 
\begin{equation}\label{modelmonte1} Y_{i}^{(1)}=   c_{i1} +  c_{i2} + \epsilon_i, \quad i=1, \ldots,n=100, \end{equation} with two regressors, and
\begin{equation}\label{modelmonte2} Y_{i}^{(2)}=   c_{i1} +  c_{i2} + c_{i3} + c_{i4}+ \epsilon_i, \quad i=1, \ldots,n=100, \end{equation}
with four regressors, both with alpha-stable \mbox{i.i.d.} $\epsilon_i$'s. The regression constants~$c_{ij}$ (the same ones across the 1000 replications) were drawn (independently) from the uniform distribution on $[-1,1]^2$ and $[-1,1]^4$, respectively. 
 Letting ${\bf 1}_K:=(1, 1, \ldots,1) \in \R^K$, the true values of the regression parameters are thus $\betab={\bf 1}_2$ in model (\ref{modelmonte1}) and $\betab={\bf 1}_4$ in model (\ref{modelmonte2}).

Denoting by ${\betab}\n(j)=({\beta}\n_1(j), \ldots ,{\beta}\n_K(j))\pr$ ($j=1,\ldots ,M$; $K=2$ or $4$ depending on the model) an estimator  ${\betab}\n $ computed from the $j$th replication,   the   empirical bias and  empirical mean square error for the first component~${\beta}\n_1 $ of~${\betab}\n $ are
$${\rm BIAS}(\betab\n):=\frac{1}{M} \sum_{j=1}^M ( {\beta}\n_1(j)- 1),
\quad  {\rm and} \quad {\rm MSE}_l(\betab\n):=\frac{1}{M} \sum_{j=1}^M ( {\beta}_1\n(j)- 1)^2,$$
respectively;  models~(\ref{modelmonte1}) and~(\ref{modelmonte2}) being perfectly symmetric,  efficiency comparisons can be based on that first component only. These quantities were computed for the least squares $\hat{\betab}\n_{\rm LS}$ and the LAD estimators~$\hat{\betab}\n_{\text{\rm LAD}}$, the one-step versions ${\utbeta}\!\!\!\phantom{^I}\n_{{J}_{\rm{vdW}}}$, ${\utbeta}\!\!\!\phantom{^I}\n_{{J}_{\rm{W}}}$ and  ${\utbeta}\!\!\!\phantom{^I}\n_{{J}_{\rm{L}}}$ of the van der Waerden, Wilcoxon and Laplace    estimators, and the one-step R-estimators  ${\utbeta}\!\!\!\phantom{^I}\n_{{J}_{\rm{\alpha/b}}}$ 
 associated with the stable scores with tail index $\alpha$ and skewness parameter $b$ ($\alpha=1.8/  b=0$; $\alpha=1.8/  b=0.5$; $\alpha=1.2/  b =0$; $\alpha=1.2/  b=0.5$; $\alpha= 0.5/ b=0.5$), respectively. For the sake of comparison, we also computed the  bias and    mean square errors associated with the Argmin (Hodges-Lehmann 1963; Jure\v ckov\' a~1971) versions ${\utbeta}\n_{\text{HL;W}}$, ${\utbeta}\n_{\text{HL;\! vdW}}$ and ${\utbeta}\n_{\text{HL;1.8/0}}$ of the Wilcoxon, van der Waerden, and stable score ($\alpha=1.8/  b=0$) R-estimators; the latter were computed via the Nelder-Mead~(1965) method. \vspace{3mm}
 
 
 Results are collected in Table~\ref{Tab1} for model~(\ref{modelmonte1}) and Table~\ref{Tab2} for model~(\ref{modelmonte2}), and confirm  the theoretical findings of the previous sections.  Least squares behave quite poorly, and fail miserably as the tail index decreases, while least absolute deviations maintain an overall good performance. The empirical performances of   R-estimators are consistent with theoretical ARE rankings.  Depending on the scores and the actual underlying tail index and skewness parameter, R-estimators may or may not improve on least absolute deviations. Stable score-based R-estimators, as a rule, outperform least absolute deviations, as expected, under correctly specified values of the tail index.  
 
 It is worth noting that one-step R-estimators are doing better than  their Hodges-Lehmann  counterparts  in tmodel (\ref{modelmonte2}), that is, when the parameter is of dimension four.  This is most probably due to computational problems related with the Argmin approach in higher dimensions; such problems do not occur in the one-step approach. Further evidence of this phenomenon is provided in Table~\ref{Tab3}, where we report results for the one-step and Hodges-Lehmann versions of the van der Waerden R-estimator  in regression models of the form
 \begin{equation}\label{modelmontek} Y_{i}^{(2)}=   c_{i1} +  c_{i2} + \ldots+ c_{iK}+ \epsilon_i, \quad i=1, \ldots,n=100, \end{equation}
with $K$ regressors, $K= 6,\ 10,\ 15$ (same number of replications; regression constants uniform over $[-1,1]^K$). Irrespective of the underlying stable density, the superiority of  the one-step version quite significantly increases with $K$.

 \begin{table}[htbp]\label{Tab1}
\begin{center}
\caption{\normalsize Empirical bias and mean square error for various  estimators of $\betab$ in model (\ref{modelmonte1})}\vspace{-0.2cm}
{\small
\begin{tabular}{lccccccc}
\hline
\hline
\vspace{0.1cm} Estimator &  \multicolumn{7}{c}{Underlying stable density $(\alpha/b)$}\\
\hline
&  &  $\alpha=2$/$b=0$ &$\alpha=1.8$/$b=0$& $\alpha=1.8$/$b=0.5$ &$\alpha=1.2$/$b=0$&$\alpha=1.2$/$b=0.5$ & $\alpha=0.5$/$\beta=0.5$  \vspace{.3 cm}\\ 
$\hat{\betab}\n_{\rm LS}$ &  ({\rm Bias}) &  .00193    & -.00134     &  .01385 & .18680   & -.19255   & 740527.6  \vspace{0mm}
\\ & ({\rm MSE}) &  .06770      &  .19459   &  .27336   & 124.46 & 88.070  & 5.3560e+14 \\ \\
$\hat{\betab}\n_{\text{\rm LAD}}$ & ({\rm Bias}) & .00167    & -.00087   & .00502  & .02995   &  .00646    & -.02438 \\ 
& ({\rm MSE}) & .10674      & .10411     &  .11638       & .11560    &  .13396    & .23233  \\ \\
${\utbeta}\!\!\!\phantom{^I}\n_{{J}_{\rm{vdW}}}$ &({\rm Bias})  & .00256    & -.00136    & .00694   & .03376  & -.00243   & .00745\\
& ({\rm MSE})& .06878   & .07694     & .08545    & .15165   & .14499   & .49418 \\ \\
${\utbeta}\!\!\!\phantom{^I}\n_{{J}_{\rm{W}}}$ &({\rm Bias}) & .00076    & .00015    & .00920 & .02957  & -.00147    & -.00165    \\
& ({\rm MSE})& .07234    & .07454    &  .08366  & .12060   & .12219   & .29830 \\ \\
${\utbeta}\!\!\!\phantom{^I}\n_{{J}_{\rm{L}}}$ &({\rm Bias})  &  .00167   & -.00087    & .00502  & .02995     & .00646  & -.02438   \\ 
& ({\rm MSE}) &  .10674  & .10411    &  .11638  & .11560    &   .13396  & .23232 \\ \\
${\utbeta}\!\!\!\phantom{^I}\n_{{J}_{\rm{1.8/0}}}$& ({\rm Bias}) & .00250    & .00063   & .00883   & .03046  & .00068  & .00267\\
& ({\rm MSE})&.07088   & .07457    & .08310     & .12976  & .12820  & .36304 \\ \\
${\utbeta}\!\!\!\phantom{^I}\n_{{J}_{\rm{1.8/.5}}}$& ({\rm Bias}) & .00187     & -.00119   & .01057   & .03284  & -.00037   & .00284 \\
& ({\rm MSE}) & .07104  & .07683    &  .08139    &  .13562  &  .12398   & .34625 \\ \\
${\utbeta}\!\!\!\phantom{^I}\n_{{J}_{\rm{1.2/0}}}$& ({\rm Bias}) & .00424   & .00353  &  .01373  & .02155    & -.00363    & .01652 \\
& ({\rm MSE}) & .11613    & .09812     &  .11040  & .09641   & .10971   & .17458 \\ \\
${\utbeta}\!\!\!\phantom{^I}\n_{{J}_{\rm{1.2/.5}}}$& ({\rm Bias}) & .00670   &  -.00418 & .01609   & .02735   & .00310    & -.00199 \\
& ({\rm MSE}) & .11416    & .10382    &  .10822  &  .11455    & .08917    & .11282  \\  \\
${\utbeta}\!\!\!\phantom{^I}\n_{{J}_{\rm{.5/.5}}}$ & ({\rm Bias}) & .01070  & .03350    & .00357    & .04768   & -.01671    & .00466 \\ 
& ({\rm MSE}) &  .22575  & .28311    & .24386    & .35926   &   .18999 & .12103 \\ \\
${\utbeta}\n_{\text{HL;\! vdW}}$& ({\rm Bias}) &  -.01668   & -.01040  &  -.00253   & .04306   & -.01664    & .11740\vspace{1mm}\\ 
& ({\rm MSE}) & .07936    & .08958    & .09508   & .20227   &  .20441   & 1.1934  \\  \\
${\utbeta}\n_{\text{HL;\! W}}$& ({\rm Bias}) & -.00672    & -.02019   &   -.01113 & -.01052   &  -.03408  & -.24449\vspace{1mm}\\ 
& ({\rm MSE})& .08225   &  .09071    &   .09702  &  .16290   &  .14918   & .82852 \\ \\
${\utbeta}\n_{\text{HL;\! 1.8/0}}$ &({\rm Bias}) &  -.02274    &  -.02834 & -.01923 &  -.01504  & -.05129      & -.24827 \vspace{1mm}\\ 
& ({\rm MSE})  & .09066       &  .10291     &  .10488   & .18247   &  .19072    & .96871  \\  \\
\hline
\hline
\multicolumn{8}{p{16cm}}{Empirical bias and MSE of the least square $\hat{\betab}\n_{\rm LS}$, the LAD   $\hat{\betab}\n_{\text{\rm LAD}}$ and various rank-based estimators computed from 1000 replications of model (\ref{modelmonte1}) with sample size $n$=100, under various stable error distributions.}\
\end{tabular}
}
\end{center} 
\end{table}

\begin{table}[htbp]\label{Tab2}
\begin{center}
\caption{\normalsize Empirical bias and mean square error for various  estimators of $\betab$ in model (\ref{modelmonte2}) }\vspace{-0.2cm}
{\small
\begin{tabular}{lccccccc}
\hline
\hline
\vspace{0.1cm} Estimator &  \multicolumn{7}{c}{Underlying stable density $(\alpha/b)$}\\
\hline
  & & $\alpha=2$/$b=0$ &$\alpha=1.8$/$b=0$& $\alpha=1.8$/$b=0.5$ &$\alpha=1.2$/$b=0$&$\alpha=1.2$/$b=0.5$ & $\alpha=0.5$/$b=0.5$  \vspace{.3 cm}\\ 
$\hat{\betab}\n_{\rm LS}$ &   ({\rm Bias}) & .00314        &  .01367   &  -.01945  & -4.09468    &  -.09272   & -47944.35  \vspace{0mm}
\\ &     ({\rm MSE}) &  .06339      &  .30161   & .12752    & 15818.91   & 39.45292  & 1.23211e+13  \\ \\
$\hat{\betab}\n_{\text{\rm LAD}}$ &  ({\rm Bias}) &  .00693      &  .00880   & -.00774    & -.00652     &  .00352  &-.00746  \\ 
&     ({\rm MSE}) &   .09995    &   .09992  & .09548   &   .08495 &  .09984   & .21871  \\ \\
${\utbeta}\!\!\!\phantom{^I}\n_{{J}_{\rm{vdW}}}$ & ({\rm Bias})  & .00378        &  .00638   & -.01177   &  -.00763   & -.01262    & -.01902 \\
& ({\rm MSE})&   .06463    &  .06964   &  .07238   &   .11369   &  .11015   & .35648  \\ \\
${\utbeta}\!\!\!\phantom{^I}\n_{{J}_{\rm{W}}}$& ({\rm Bias}) &  .00542      & .00579    &  -.01236  & -.00624     &  -.00774   & -.01330    \\
& ({\rm MSE})&    .06811   & .06847    & .06988    & .09038     & .09127     & .22657 \\ \\
${\utbeta}\!\!\!\phantom{^I}\n_{{J}_{\rm{L}}}$& ({\rm Bias})  &  .00693     &  .00880    &    -.00774   &  -.00652    & .00352     &  -.00746  \\ 
&({\rm MSE})&   .09995    &   .09992  & .09548    & .08495   &  .09984   & .21871 \\ \\
${\utbeta}\!\!\!\phantom{^I}\n_{{J}_{\rm{1.8/0}}}$& ({\rm Bias}) &  .00499     &  .00531   &  -.01221  & -.00445    & -.00980     & -.01629 \\
& ({\rm MSE})&   .06755    &  .06735   &  .07021    & .09908     &   .09562   & .27044 \\ \\
${\utbeta}\!\!\!\phantom{^I}\n_{{J}_{\rm{1.8/.5}}}$& ({\rm Bias}) &  .00339        &   .00526  & -.01109   &  -.00438  & -.01151     & -.01722  \\
& ({\rm MSE})& .06686     & .06914     &  .06977    & .10095     &   .09397  & .25358 \\ \\
${\utbeta}\!\!\!\phantom{^I}\n_{{J}_{\rm{1.2/0}}}$  &({\rm Bias}) &  .00802      &   .00608  &  -.01297   &  .00682   & .00404     & .00226  \\
& ({\rm MSE})&  .10763     &  .09229   & .08986    & .07061   & .08406   & .13542  \\ \\
${\utbeta}\!\!\!\phantom{^I}\n_{{J}_{\rm{1.2/.5}}}$& ({\rm Bias}) &  .00291      &  .00024    &  -.01401  & .00396     & -.00231     & -.00573 \\
& ({\rm MSE})&  .10332    &  .09233   &   .08567  & .09036    &    .07037  & .07636  \\  \\
${\utbeta}\!\!\!\phantom{^I}\n_{{J}_{\rm{.5/.5}}}$ & ({\rm Bias}) & .03400      &  .03653   & -.02823   &  -.05925 
   & -.00469     & -.01970  \\ 
& ({\rm MSE})&  .30150     &  .35030    & .28818    & .43049    &  .18807  &  .19423 \\ \\
${\utbeta}\n_{\text{HL;\! vdW}}$ &({\rm Bias}) & .00401       & .00634   & -.01208   & -.00704   & -.01234    & -.02138  \vspace{1mm}\\ 
&  ({\rm MSE})&  .06513      &   .06968   &  .07266   & .11310    &  .10956   & .38167  \\  \\
${\utbeta}\n_{\text{HL;\! W}}$& ({\rm Bias}) & .00513        &  .00623    & -.01285  &   -.00547 &  -.00755   &  -.01470 \vspace{1mm}\\ 
& ({\rm MSE}) & .06854      & .06855     & .07006    & .09010   &  0.09100   &  .23734 \\ \\
${\utbeta}\n_{\text{HL;\! 1.8/0}}$& ({\rm Bias}) &  .00494      &  .00582   & -.01245   & -.00396    &  -.01081   & -.01793  \vspace{1mm}\\ 
& ({\rm MSE})  & .06783      &  .06753   & .07037    & .09854     & .09594   & .28729  \\  \\
\hline
\hline
\multicolumn{8}{p{16cm}}{Empirical bias and MSE of the least square $\hat{\betab}\n_{\rm LS}$, the LAD   $\hat{\betab}\n_{\text{\rm LAD}}$ and various rank-based estimators computed from 1000 replications of model (\ref{modelmonte2}) with sample size $n$=100, under various stable error distributions.}\
\end{tabular}
}
\end{center} 
\end{table}

 \begin{table}[htbp]
\begin{center}
\caption{\normalsize One-step R-estimation versus Argmin}\vspace{-0.2cm}
{\small
\begin{tabular}{lccccccc}
\hline
\hline
\vspace{0.1cm} Estimator &  \multicolumn{7}{c}{Underlying stable density $(\alpha/b)$}\\
\hline   
 &&  $\alpha=2$/$b=0$ &$\alpha=1.8$/$b=0$& $\alpha=1.8$/$b=0.5$ &$\alpha=1.2$/$b=0$&$\alpha=1.2$/$b=0.5$ & $\alpha=0.5$/$b=0.5$ 
 \\  
 $K=6$\vspace{-2mm}&&&&&&&\\  \\ 
${\utbeta}\!\!\!\phantom{^I}\n_{{J}_{\rm{vdW}}}$ &$\quad$  ({\rm Bias}) & -.01991   & -.00485   &  .01084   &  -.01890   & .02246    & .00162   \vspace{0mm}
\\ &$\quad$  ({\rm MSE})  & .07707   &  .08821    & .08935   & .16485   &  .15258  & .61554\vspace{2mm} \\ 
${\utbeta}\n_{\text{HL;\! vdW}}$ &$\quad$  ({\rm Bias})  & -.19519    & -.19834   & -.19202   & -.36809   &  -.30435 & -.59222  \vspace{0mm}
\\  &$\quad$   ({\rm MSE})  & .24257    &  .27483   &   .27461 & .58981  & .52245    & 2.51344 \\ \\
\hline \\
 $K=10$\vspace{-2mm}&&&&&&&\\  \\  
${\utbeta}\!\!\!\phantom{^I}\n_{{J}_{\rm{vdW}}}$ &$\quad$  ({\rm Bias})  & -.00877    &  .00607   & .00187    & -.00807   & -.01376     & .06003    \vspace{0mm}
\\ &$\quad$   ({\rm MSE})  &  .07834  &  .09133   & .08641    & .16835  &  .15545  & 1.4346\vspace{2mm}  \\ 
${\utbeta}\n_{\text{HL;\! vdW}}$& $\quad$  ({\rm Bias})  & -.91080   &   -.89626  & -.92196  & -1.00979  & -.99976   &   -.97662  \vspace{0mm}
\\  &$\quad$   ({\rm MSE})  & 1.04321    & 1.07289    & 1.09949   & 1.50269   & 1.43327   & 3.23870  \\ \\
\hline \\  $K=15$ \vspace{-2mm}& & & & & & &\\  \\  
${\utbeta}\!\!\!\phantom{^I}\n_{{J}_{\rm{vdW}}}$& $\quad$  ({\rm Bias})  & -.00374   & -.01421     &  -.00575   & .02479  & 0.00271    &   .01123 \vspace{0mm}
\\ & $\quad$   ({\rm MSE}) & .08894   & .10969   &  .10539    &  .20918  &  .19621   & 2.00335 \vspace{2mm} \\  
${\utbeta}\n_{\text{HL;\! vdW}}$ &$\quad$  ({\rm Bias})  & -1.07573   &  -1.11915   & -1.11057   & -1.23107  & -1.21492   & -1.31910 \vspace{0mm}
\\ & $\quad$   ({\rm MSE})  & 1.19685    &  1.33319   &  1.32890 & 1.91879   &  1.88120   & 4.32374 \vspace{3mm}\\
\hline
\hline
\multicolumn{8}{p{17cm}}{Empirical bias and MSE of the one-step and Argmin versions ${\utbeta}\!\!\!\phantom{^I}\n_{{J}_{\rm{vdW}}}$ and ${\utbeta}\n_{\text{HL;\! vdW}}$ of the van der Waerden R-estimator  computed from 1000 replications of model (\ref{modelmontek}) with $K= $ 6, 10, 15, sample size $n$=100 and various stable error distributions.}\
\end{tabular}
}
\end{center} 
\label{Tab3}
\end{table}

\section{Conclusion.}

Stable densities constitute    a broad and   flexible class of probability density functions, allowing for asymmetry and heavy tails. Their theoretical properties make them   quite appealing in a variety of applications, including econometric and financial ones. Traditional inference methods, however, in general are not valid    in models involving stable error: classical tests no longer satisfy  nominal probability level  constraints, and estimators, as a rule, are rate-suboptimal. On the other hand, due to the absence of closed-form likelihoods,  theoretical optimality results are not easily derived.  And, still for the same reason,  their practical implementation is all but straightforward.

In the particular case of linear models with stable errors (with unspecified tail index $\alpha$ and skewness parameter $b$), Hallin \emph{et al.} (2010) show how rank-based methods provide a powerful and convenient solution to testing problems. In order to do so, they first establish the local asymptotically normal nature (ULAN, with root-$n$ contiguity rates)  of linear model experiments with stable errors. In this paper, we extend their approach to estimation problems. More particularly, taking full advantage of the ULAN property, we construct one-step R-estimators for the regression parameter $\betab$. Those estimators are root-$n$ consistent  and asymptotically normal, irrespective of the underlying stable density, and  their asymptotic covariance matrices are obtained  as a by-product of the one-step procedure. Using numerical results derived in Hallin~\emph{et~al.}~(2010),   we  moreover show how to construct the R-estimators  associated with stable scores, achieving parametric optimality at prespecified values of $\alpha$ and $b$.  

A  thorough Monte Carlo study    confirms   the excellent  finite-sample  performances of  our one-step\linebreak  R-estimators, which are shown to outperform  not only the traditional OLS and LAD estimator, but also their Argmin or Hodges-Lehmann counterparts.

\section{Appendix.}

 \noindent {\bf Proof of Proposition \ref{asymplin}.}
 Point (i) is a direct consequence of  the H\'ajek projection theorem.  Points (ii) and (iii) follow from point (i), the central limit theorem and the Le Cam's Third Lemma. As for point (iv), Theorem 3.1 in Jure\v ckov\'a (1969) applies.\cqfd \vspace{4mm}

 \noindent {\bf Proof of Proposition \ref{asympnorm}.} In view of (\ref{ourestimate}), we have that
 \begin{equation} \label{a1} n^{1/2}({\utbeta}\!\!\!\phantom{^I}\n_{J}- \betab)=n^{1/2}(\hat{\betab}\phantom{i}\!\!\n_{\scriptscriptstyle{\text{\rm LAD}}}-\betab)+ \mathbb{K}\n \widehat{\mathcal{J}}^{-1}(J, g) \utDelta_{J}^{(n)}(\hat{\betab}\phantom{i}\!\!\n_{\scriptscriptstyle{\text{\rm LAD}}}). \end{equation}
The consistency of $\widehat{\mathcal{J}}^{-1}(J, g)$ together with point (iv) of Proposition \ref{asymplin} entail that, under $ {\rm P}\n_{g, a , \betab}$ with $g \in \mathcal{F}$, as $\ny$,
\begin{eqnarray} \label{a2} \mathbb{K}\n \widehat{\mathcal{J}}^{-1}(J, g) \utDelta_{J}^{(n)}(\hat{\betab}\phantom{i}\!\!\n_{\scriptscriptstyle{\text{\rm LAD}}})&=& \mathbb{K}\n {\mathcal{J}}^{-1}(J, g) \utDelta_{J}^{(n)}({\betab})- n^{1/2}(\hat{\betab}\phantom{i}\!\!\n_{\scriptscriptstyle{\text{\rm LAD}}}- \betab) + o_{\rm P}(1).\end{eqnarray} 
Combining (\ref{a1}) and (\ref{a2}), we readily obtain 
 \begin{equation} \label{eqend} n^{1/2}({\utbeta}\!\!\!\phantom{^I}\n_{J}- \betab)=\mathbb{K}\n {\mathcal{J}}^{-1}(J, g) \utDelta_{J}^{(n)}({\betab})+ o_{\rm P}(1) \end{equation}
under $ {\rm P}\n_{g, a , \betab}$ with $g \in \mathcal{F}$, as $\ny$. The result follows using Proposition \ref{asymplin}. \cqfd \vspace{4mm}

   \noindent {\bf Proof of Proposition  \ref{LADvsLap}. } Without loss of generality, we assume that the $\epsilon_i$'s have median zero. In this proof, we show that $n^{1/2} (\hat{\betab}\n_{\scriptscriptstyle{\text{\rm LAD}}}- {\betab})=n^{1/2}({\utbeta}\!\!\!\phantom{^I}\n_{{J}_{\text{L}}}- \betab)+o_{\rm P}(1)$. From the proof of Theorem 4.1 in Koenker~(2005) (see also Koenker and Basset~1978), we have that (least absolute deviation estimation is equivalent to median regression hence quantile regression   with quantile of order  $\tau=1/2$)
 \begin{equation} \label{eqLAD} 
n^{1/2} (\hat{\betab}\n_{\scriptscriptstyle{\text{\rm LAD}}}- {\betab})= \frac{n^{-1/2}}{2 g(0)} \mathbb{K}\n\mathbb{K}^{(n)\prime} \sum_{i=1}^n {\rm sign} (\epsilon_i) c_i + o_{\rm P}(1) \end{equation}
 under ${\rm P}\n_{g, a , \betab}$. Now, since $J_{\rm L}(u)=\sqrt{2} {\rm sign}(u-1/2)$, we have that  
 \begin{eqnarray} \label{crossLa} 
 \mathcal{J}(J_{\rm L},g) 
 & = &\sqrt{2} 
 \int_{0}^1 {\rm sign}(u-1/2) \varphi_g(G^{-1}(u)) du \nonumber \\ 
 &=& -\sqrt{2} \int_{- \infty}^{\infty} {\rm sign}(G(v)-G(0))  g\pr(v)dv \nonumber \\ 
 &=& \sqrt{2} \int_{-\infty}^0 g\pr(v)dv - \sqrt{2} \int_0^{\infty} g\pr(v) dv 
 = 2 \sqrt{2} g(0).
 \end{eqnarray} 
 Using (\ref{crossLa}), (\ref{eqend}) in the proof of Proposition \ref{asympnorm} and point (i) of Proposition \ref{asymplin}, we obtain that 
\begin{eqnarray*} 
 n^{1/2}({\utbeta}\!\!\!\phantom{^I}\n_{J_{\text{L}}}- \betab) 
 &= &
 \mathbb{K}\n {\mathcal{J}}^{-1}({J_{\text{L}}}, g) \utDelta_{J_{\text{L}}}^{(n)}({\betab})+ o_{\rm P}(1) \\ 
 &=& \mathbb{K}\n {\mathcal{J}}^{-1}({J_{\text{L}}}, g) \Deltab_{J_{\text{L}}}^{(n)}({\betab})+ o_{\rm P}(1) \\
 &=& \frac{n^{-1/2}}{2 g(0)} \mathbb{K}\n\mathbb{K}^{(n)\prime} \sum_{i=1}^n {\rm sign} \left(G(\epsilon_i)-\frac{1}{2}\right) c_i + o_{\rm P}(1) \\
  &=& \frac{n^{-1/2}}{2 g(0)} \mathbb{K}\n \mathbb{K}^{(n)\prime} \sum_{i=1}^n {\rm sign} \left(\epsilon_i \right) c_i + o_{\rm P}(1), 
  \end{eqnarray*}
 which, in view of (\ref{eqLAD}),   completes the proof.\cqfd  \vspace{3mm}

\noindent{\bf Acknowledgements. }
 The authors thank two anonymous referees for their pertinent comments on their original manuscript, which led to a much improved revised version. 

Marc Hallin acknowledges the support of the Sonderforschungsbereich
``Statistical modelling of non-linear dynamic processesÓ (SFB 823)
of the Deutsche Forschungsgemeinschaft, and a Discovery Grant of the
Australian Research Council. The research of Yvik Swan was partially
supported by a Mandat de Charg\' e de Recherche from the Fonds
National de la Recherche Scientifique, Communaut\' e fran\c caise de
Belgique, and that of Thomas Verdebout by a BQR (Bonus Qualit\' e
Recherche) of the Universit\' e  Lille~3. David Veredas acknowledges
the financial support of the IAP P6/07 contract, from the IAP program
(Belgian Federal Scientific Policy) ``Economic policy and finance in
the global economy".

\vspace{3mm}


\begin{thebibliography}{plainnat}

\bibitem{[1]} Allal, J., A. Kaaouachi and D. Paindaveine, 2001, R-estimation for ARMA models.  Journal of Nonparametric Statistics 13, 815-831.

\bibitem{[2]} Andrews, B., M. Calder and R. A. Davis, 2009,  Maximum likelihood estimation for $\alpha$-stable autoregressive processes.  The Annals of  Statistics  37, 1946-1982.

\bibitem{[3]}  Bassett, G. and R. Koenker, 1978, Asymptotic theory of least absolute error regression. Journal of the American Statististical Association 73,  618-622.

\bibitem{[4]}  Blattberg, R. and T. Sargent, 1971, Regression with Non-Gaussian Disturbances: Some Sampling Results. Econometrica 39, 501-510.

\bibitem{[5]} Cassart, D., M. Hallin and D. Paindaveine, 2010,  On the estimation of cross-information quantities in R-estimation,  in: J. Antoch, M. Huskov\' a and P.K. Sen, (Eds.), Nonparametrics and Robustness in Modern Statistical Inference and Time Series Analysis: A Festschrift in Honor of Professor Jana Jure\v ckov\' a, I.M.S. Collections, Volume 7, 35-45.

\bibitem{[51]} Deo, R.S., 2002, On testing the adequacy of stable processes under conditional heteroskedasticity. Journal of Empirical Finance 9, 257Ð270. 

\bibitem{[52]} Dominicy, Y.  and D. Veredas, 2010, The method of simulated quantiles,  ECARES WP 2010/09.

\bibitem{[6]}  DuMouchel, W. H., 1973, On the asymptotic normality of the maximum-likelihood estimate when sampling from a stable distribution. The Annals of  Statistics 1,  948-957. 

\bibitem{[7]} DuMouchel, W. H., 1975, Stable distributions in statistical inference . 2: Information from stably distributed samples. Journal of the American Statistical Association 70, 386-393. 

\bibitem{[8]} El Bantli, F. and  M. Hallin, 1999,  $L_1$-estimation in linear models with heterogeneous white noise. Statistics and Probability Letters 45,  305-315.

\bibitem{[9]} El Barmi, H. and P.I. Nelson, 1997,  Inference from stable distributions. Proceedings of the Symposium on Estimating Functions. IMS Lecture Notes-Monograph Series,  32, 439-456.

\bibitem{[10]} Fama, E. and R. Roll, 1971, Parameter Estimates for Symmetric Stable Distributions. Journal of the American Statistical Association  66, 331-338.

\bibitem{[11]} Hallin, M.,   H. Oja   and D. Paindaveine, 2006, Semiparametrically efficient rank-based inference for shape. II. Optimal R-estimation of shape. The Annals of Statistics 34, 2757-2789. 

\bibitem{[12]} Hallin, M. and D. Paindaveine, 2008, Semiparametrically efficient one-step R-Estimation. Unpublished manuscript.

\bibitem{[13]} {Hallin}, M.,  Y. Swan, T. Verdebout and D. Veredas, 2010,  Rank-based testing in linear models with stable errors.  Journal of Nonparametric Statistics, to appear. 

\bibitem{[14]}  Hodges, J. L. and E. L. Lehmann,  1963, Estimation of location based on ranks. The Annals of Mathematical Statistics 34, 598-611.

\bibitem{[15]} Jure\v{c}kov\'a, J., 1969, Asymptotic linearity of a rank statistic in regression parameter. The Annals of Mathematical Statistics 40, 1889-1900.

\bibitem{[16]} Jure\v{c}kov\'a, J., 1971, Nonparametric estimate of regression coefficients. The Annals of Mathematical Statistics 42, 1328-1338. 

\bibitem{[17]} Jure\v{c}kov\'a, J. and P. K. Sen, 1996, Robust Statistical Procedures: Asymptotics and Interrelations, Wiley, New York. 

\bibitem{[18]} Knight, K., 1998, Bootstrapping sample quantiles in non-regular cases. Statistics and Probability Letters 37, 259-267. 

\bibitem{[19]} Koenker, R., 2005, Quantile Regression, Cambridge University Press, Cambridge. 

\bibitem{[20]} {Koul}, H. L., 1971, Asymptotic behavior of a class of confidence regions based on ranks in regression. Annals of Mathematical Statistics 42, 466-476.

\bibitem{[21]} {Koul}, H. L., 2002,  Weighted empirical processes in dynamic nonlinear models, Lecture Notes in Statistics 166, Springer, New York. 

\bibitem{[22]} {Koul}, H. L. and  A.   Saleh,  1993, R-estimation of the parameters of autoregressive [AR (p)] models. The Annals of Statistics 21, 534-551.

\bibitem{[24]}  Le Cam, L. and  Yang, G. L. (2000), Asymptotics in
  Statistics, 2nd edition, Springer-Verlag, New-York.


\bibitem{[24]} Lehmann, E. L., 1963, Nonparametric confidence intervals for a shift parameter. The Annals of Mathematical Statistics 34, 1507-1512.

\bibitem{[25]}  Mittnik, S., M.S. Paolella and S.T. Rachev, 2000, Diagnosing and treating the fat tails in financial returns dat.  Journal of Empirical Finance 7, 389Ð416.

\bibitem{[25b]} Nelder, J.A. and Mead, R., 1965, A Simplex Method for Function Minimization. Computer Journal 7 (4): 308-313.

\bibitem{[26]} Nolan, J. P., 1997, Numerical calculation of stable densities and distribution functions. Communications in Statistics and  Stochastic Models 13,  759-774.

\bibitem{[27]} Nolan, J. P., 1999, An algorithm for evaluating stable densities in Zolotarev's (M) parametrization. Mathematical and Computer Modelling  29,  229-233.

\bibitem{[28]} Samorodnitsky, G., S.T. Rachev, J.R. Kurz-Kim, and S.V. Stoyanov, 2007, Asymptotic distributions of unbiased linear estimators in the presence of heavy-tailed stochastic regressors and residuals. Probability and Mathematical Statistics 27, 275-302. 

\bibitem{[29]} Samorodnitsky, G., and M. S. Taqqu, 1994, Stable Non-Gaussian Random Processes: Stochastic Models with Infinite Variance, Chapman and Hall, New York.

\bibitem{[30]} Sen, P. K., 1966, On a distribution-free method of estimating asymptotic efficiency of a class of nonparametric tests. Annals of Mathematical Statistics 37, 1759-1770. 

\bibitem{[31]} Wise, J., 1966, Linear estimators for linear regression
systems having infinite residual variances.  Unpublished manuscript, Department
of Economics, University of Hawaii.

\bibitem{[32]} Zolotarev, V. M., 1986, One-dimensional stable distributions,   Translation of Mathematical Monographs 65, American  Mathematical  Society, Providence, RI.

\bibitem{[33]} Zolotarev, V. M., 1995, On representation of densities of stable laws by special functions. Theory of Probability and its Applications  39, 354-362.

\end{thebibliography}
\end{document}